\documentstyle[11pt,newpasp,twoside,epsf]{article}
\def\edcomment#1{\iffalse\marginpar{\raggedright\sl#1\/}\else\relax\fi}
\reversemarginpar

\begin{document}
\title{Ionized Iron Lines in X-ray Reflection Spectra}
\author{D.R. Ballantyne and A.C. Fabian}
\affil{Institute of Astronomy, University of Cambridge, Cambridge,
United Kingdom, CB3 OHA}
\author{R.R. Ross}
\affil{Physics Department, College of the Holy Cross, Worcester, MA
01610, USA}

\begin{abstract}
We present results from new calculations of the X-ray reflection
spectrum from ionized accretion discs. These computations improve on
our previous models by including the condition of hydrostatic balance
in the vertical direction, following the work of Nayakshin, Kazanas \&
Kallman. We find that an ionized Fe K$\alpha$ line is prominent in the
reflection spectra for a wide variety of physical conditions. The
results hold for both gas and radiation pressure dominated discs and
when the metal abundances have been varied.  
\end{abstract}

\section{Introduction}
The iron K$\alpha$ line is arguably the most important feature in the
hard X-ray spectrum of Active Galactic Nuclei (AGN) and Galactic
Black Hole Candidates (GBHCs). Sensitive measurements of the shape and
energy of the line could provide information on the velocity and
ionization state of the accretion flow, as well as constraining the
properties of the central black hole (see the recent review by Fabian
et al.\ (2000) and references therein). The line is thought to arise
from the fluorescence of irradiated optically-thick material, most
likely the accretion disc itself. With the launch of the
high-throughput telescope {\it XMM-Newton}, high signal-to-noise
detections of the Fe K$\alpha$ line will soon be common. It is
therefore important that the models of X-ray reflection are suitably
advanced to take advantage of this high-quality data.

In this contribution, we present new calculations of X-ray reflection from an
ionized accretion disc, and concentrate on the properties of the Fe
K$\alpha$ line. It will be shown that emission from ionized Fe is
quite common over a wide range of physical conditions. 

\section{Computations}
The calculations were performed using the code described in detail by
Ross \& Fabian (1993) and Ballantyne, Ross, \& Fabian (2001). The
program computes the reflection spectrum from an optically thick layer
of gas over the energy range 0.001 to 100~keV. The incident radiation
is in the form of a power-law with photon index $\Gamma$ which
strikes the gas with a flux $F_x$ and at an incidence angle of $i$
degrees to the normal. The atmosphere then relaxes into thermal, ionization and pressure equilibrium
before the reflection spectrum is computed. The transfer of the
illuminating radiation is treated analytically in a one-stream
approximation. The diffuse radiation (that from the disc, the gas, and
the scattered component of the illuminating radiation) is treated
using the Fokker-Planck/diffusion method (Ross, Weaver, \& McCray
1978). The calculations are one-dimensional and occur at a radius $r$
along the accretion disc ($r=9$ Schwarszchild radii for all the models
presented below). We include levels from C~{\sc v--vii}, O~{\sc v--ix}, Mg~{\sc
ix--xiii}, Si~{\sc xi--xv}, and Fe~{\sc xvi--xxvii} with the
abundances of Morrison \& McCammon (1983). The radiation
pressure from both the impinging and diffuse radiation is included in
determining hydrostatic balance. Calculations were performed using
both the gas and radiation pressure dominated boundary conditions with
no energy dissipation in the corona as described by Merloni, Fabian,
\& Ross (2000). 

\section{Results}
Figure~1 presents results from the models run with the assumption of a
gas pressure dominated accretion disc.
\begin{figure}[ht!]
\plottwo{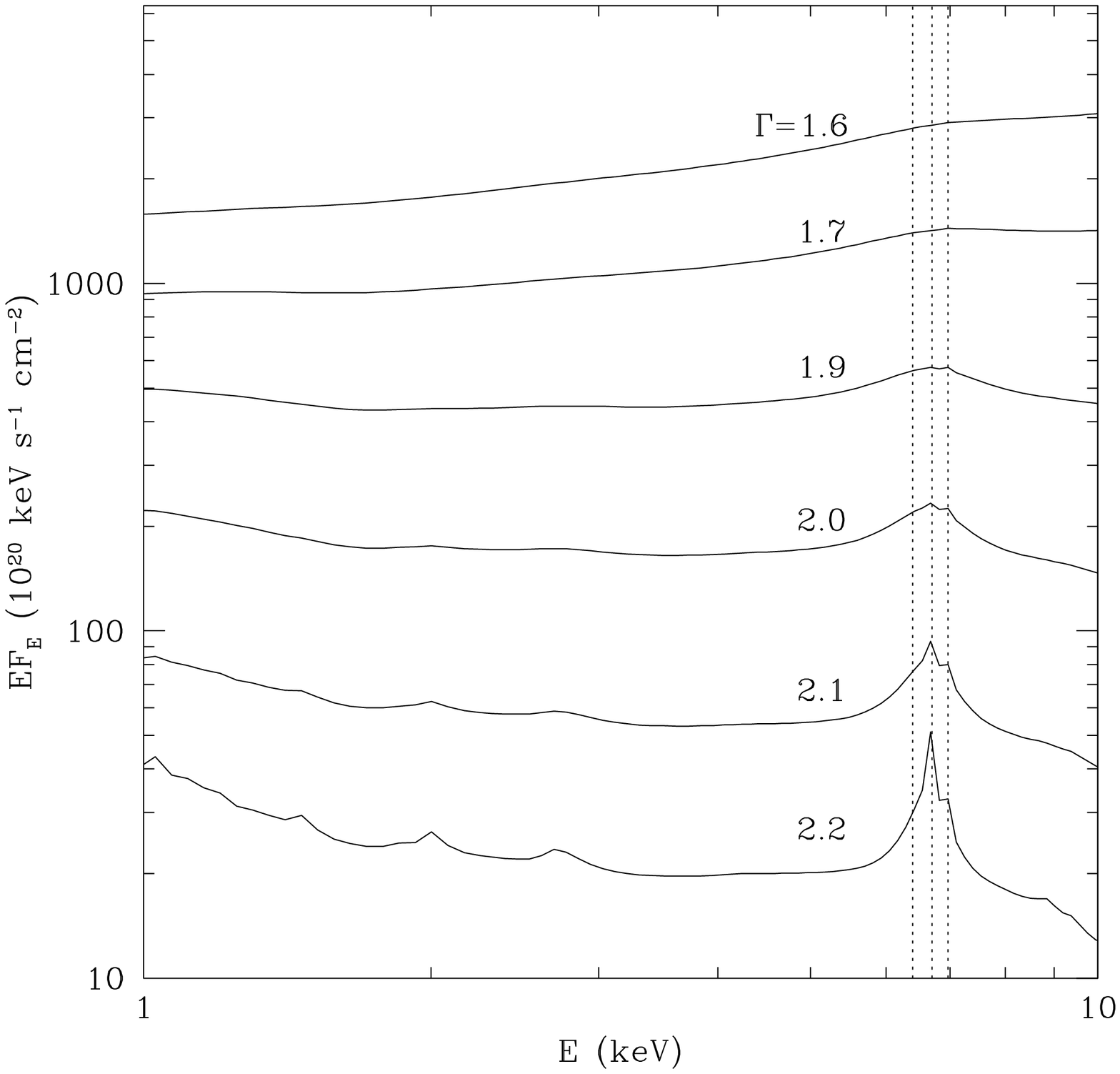}{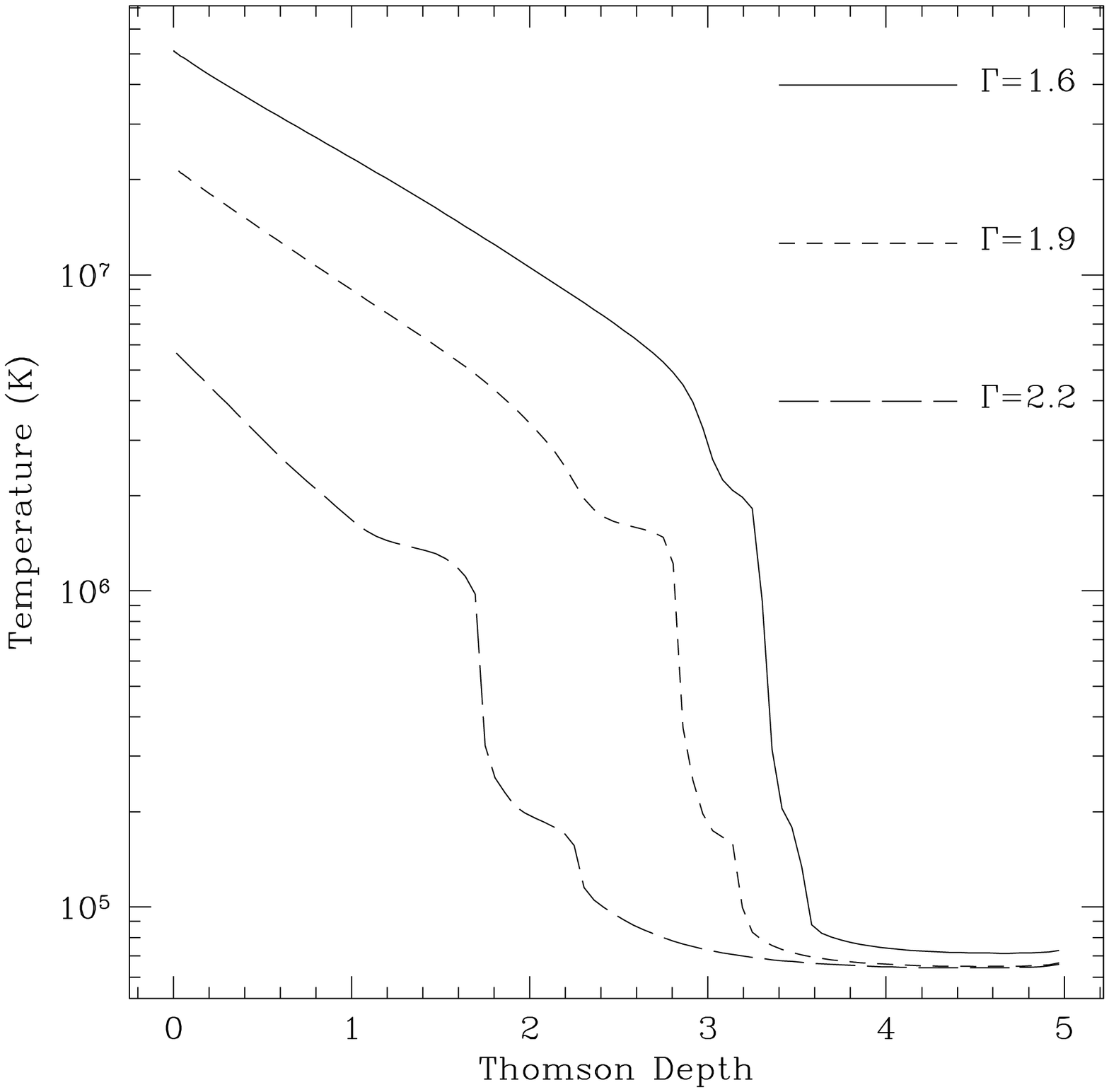}
\caption{\emph{(Left)} X-ray reflection spectra, plotted between 1 and 10~keV
to emphasize the Fe~K$\alpha$ line, for differing values of the
illuminating photon index, $\Gamma$. These models were calculated
assuming the disc was gas pressure dominated with an accretion rate of
$\dot m = 0.001$ for a 10$^{8}$~M$_{\sun}$ black hole. The spectra
have been offset vertically for clarity. The dotted vertical lines
show the positions of the iron lines at 6.4, 6.7 and 6.97~keV. \emph{(Right)} The
temperature of the atmosphere as a function of Thomson depth for three
different values of $\Gamma$. The shoulder at $T \sim 2 \times 10^6$~K
is stable due to a balance of photoelectric heating and a combination
of line and bremsstrahlung cooling.}
\end{figure}
In these cases, the illuminating radiation was incident on a disc that
was accreting at 0.001 of the Eddington rate for a 10$^8$~M$_{\sun}$
black hole ($\dot m = 0.001$). The incident flux was 10$^{15}$~erg~cm$^{-2}$~s$^{-1}$,
the incidence angle was set so that $\cos i = 1/\sqrt{3}$, and
$\Gamma$ was varied from 1.6 to 2.2. The left hand panel shows that
ionized Fe lines at 6.7 and 6.97~keV were found to be prominent for larger values of
$\Gamma$. The reason for this is illustrated in the right
hand panel of Figure~1. Here, we plot the temperature of the gas
versus the Thomson depth of the gas in the illuminated atmosphere. The
gas temperature decreases rapidly, but not discontinuously
(cf. Nayakshin, Kazanas, \& Kallman 2000). There is a zone at $
\sim 2\times 10^6$~K which is kept thermally stable by a balance of
photoelectric heating and a combination of line and bremsstrahlung
cooling. This zone of stability allows sufficient quantities of
helium-like and hydrogenic iron to exist and imprint their
features on the reflection spectrum. For harder illuminating spectra
(smaller $\Gamma$), iron is fully ionized to greater depth until the Fe lines disappear almost entirely. Ballantyne et al.\ (2001) also find
ionized Fe lines when the other model parameters, such as $F_x$, have
been varied.   

Figure 2 presents Fe lines that were calculated for two other physical situations.
\begin{figure}[hbt!]
\plottwo{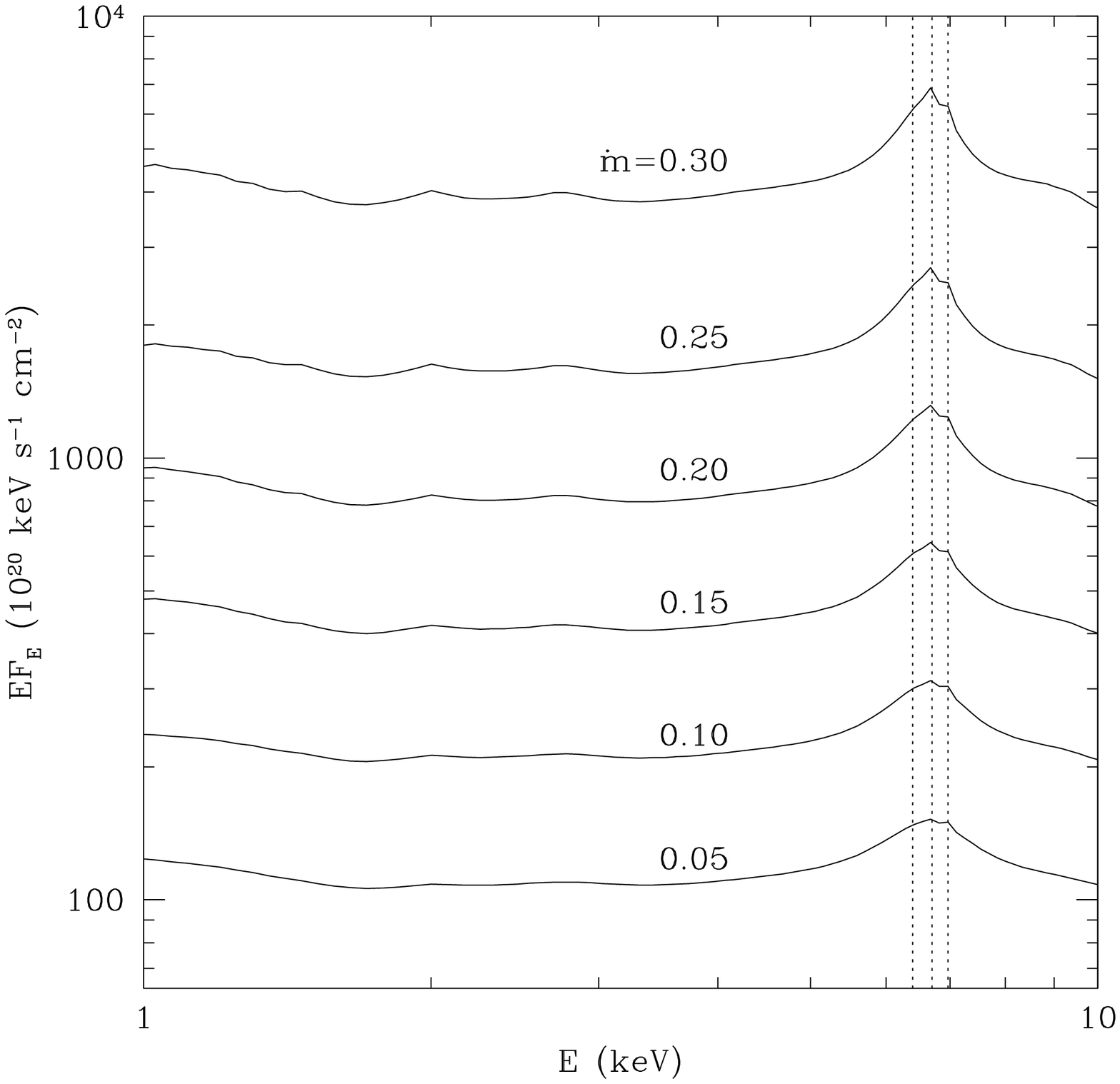}{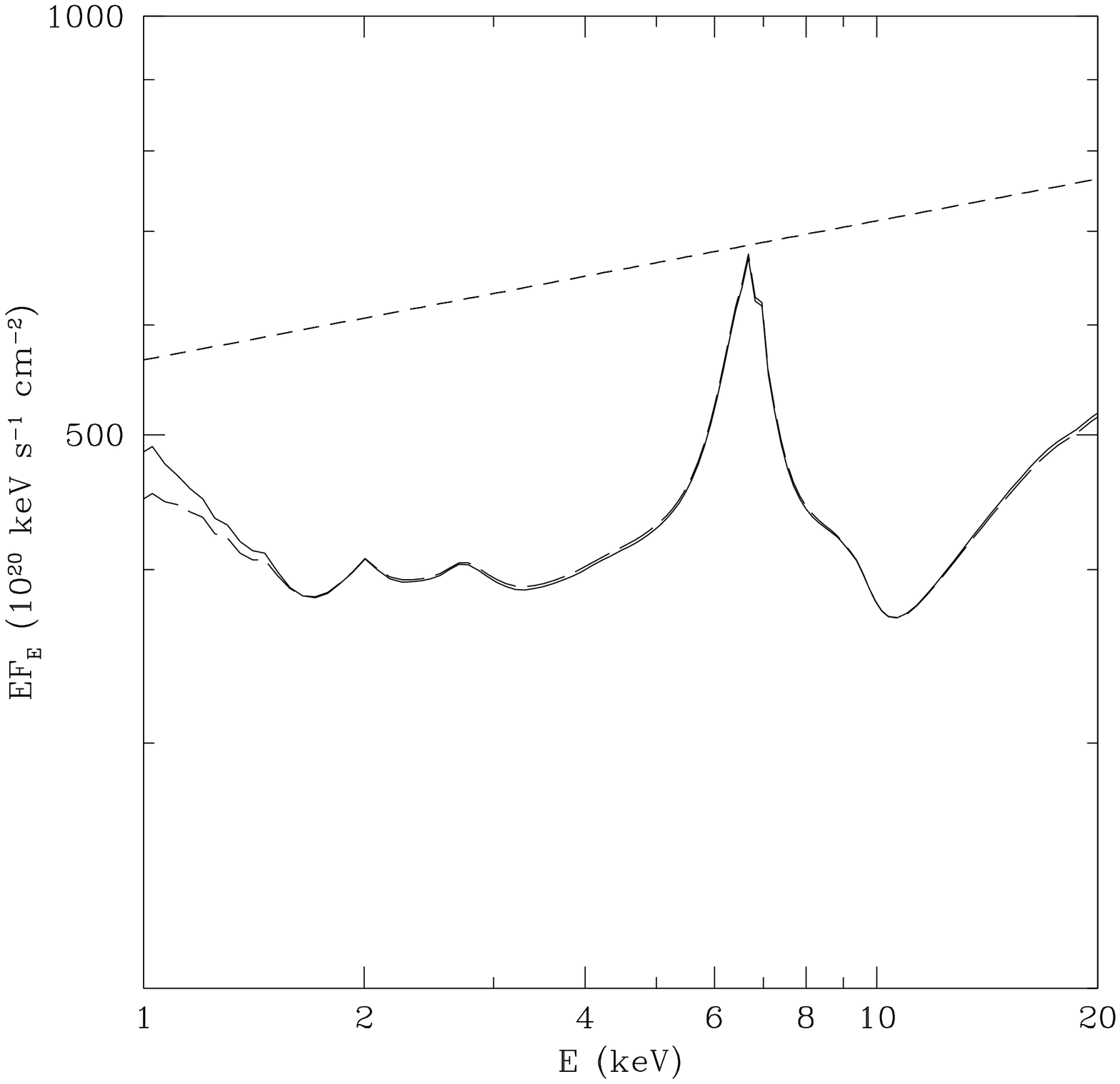}
\caption{\emph{(Left)} X-ray reflection spectra, plotted between 1 and 10~keV
to emphasize the Fe~K$\alpha$ line, for differing values of the
dimensionless accretion rate $\dot m$. The spectra have been offset
vertically for clarity. These models were calculated assuming a
radiation pressure dominated accretion disc around a black hole of
mass 10$^8$~M$_{\sun}$. The incident power-law had the following
parameters: $\Gamma=1.9$, $F_x=10^{15}$~erg~cm$^{-2}$~s$^{-1}$, and
$\cos i= 1/\sqrt{3}$. As in Fig.~1, the dotted lines denote the
position of the three different Fe lines. \emph{(Right)}. A comparison of
models with one-fifth (solid line) and solar (dashed) abundance of
oxygen. The straight dashed line denotes the continuum: $\Gamma=1.9$,
$F_x=10^{15}$~erg~cm$^{-2}$~s$^{-1}$, and $\cos i=1/\sqrt{3}$. The
accretion rate was $\dot m =0.25$ for both these models and radiation
pressure dominated boundary conditions were assumed.}
\end{figure}
The left-hand panel presents results for when radiation pressure
dominated boundary conditions were assumed, allowing calculations at
higher accretion rates. In these models, the accretion rate around a
10$^8$~M$_{\sun}$ black hole was varied, and the illuminating
parameters were held fixed at the following values: $\Gamma=1.9$,
$F_x=10^{15}$~erg~cm$^{-2}$~s$^{-1}$, and $\cos i=1/\sqrt{3}$. As
before, a line at 6.7~keV from helium-like Fe was found for every
value of the accretion rate that was considered. The right-hand panel
of Figure~2 compares two different calculations of the $\dot m =0.25$
model from the left-hand panel. The solid line denotes a model that was
calculated with an oxygen abundance one-fifth of solar, while the
dashed line denotes the solar abundance model from the other
panel. Changing the abundance of oxygen could potentially impact the
stability of the small plateau in the temperature structure of the
atmosphere, and therefore affect which Fe K$\alpha$ line dominates the
reflection spectrum. However, as Fig. 2 illustrates, changing the
oxygen abundance does not have an impact on the ionization state of
the predicted Fe K$\alpha$ line.

These results show that ionized Fe lines are expected from AGN
reflection spectra over a wide range of physical conditions. This
could have important implications on determining the structure of
accretion discs and how they are illuminated. However, disentangling
ionization effects from instrumental and/or relativistic ones will be
a challenge with current instruments. 

\acknowledgements

DRB is grateful for financial support from the Commonwealth
Scholarship and Fellowship Plan and the Natural Sciences and
Engineering Research Council of Canada. RRR and ACF acknowledge
financial support from the College of the Holy Cross and the Royal
Society, respectively.

\end{document}